\documentclass{aa}  

\usepackage{graphicx}
\usepackage{txfonts}
\usepackage{natbib}
\bibpunct{(}{)}{;}{a}{}{,} 
\begin{document}

   \title{Water vapor as a probe of the origin of gas in debris disks}


   \author{Yasuhiro Hasegawa
          \inst{1}
          \and
          Riouhei Nakatani
          \inst{1,2}
          \and
          Isabel Rebollido
          \inst{3}
          \and
          Meredith MacGregor
          \inst{4}
          \and
          Bj\"{o}rn J. R. Davidsson
          \inst{1}
          \and
          Dariusz C. Lis
          \inst{1}
          \and
          Neal Turner
          \inst{1}
          \and
          Karen Willacy
          \inst{1}
          }

   \institute{Jet Propulsion Laboratory, California Institute of Technology, Pasadena, CA 91109, USA\\
              \email{yasuhiro.hasegawa@jpl.nasa.gov}
         \and
             RIKEN Cluster for Pioneering Research, 2-1 Hirosawa, Wako-shi, Saitama 351-0198, Japan
          \and
             European Space Agency (ESA), European Space Astronomy Centre (ESAC), Camino Bajo del Castillo s/n, 28692 Villanueva de la Cañada, Madrid, Spain
          \and
             Department of Physics and Astronomy, Johns Hopkins University, 3400 N Charles Street, Baltimore, MD 21218, USA   
             }

   \date{Received; accepted}

 
  \abstract
   {Debris disks embrace the formation and evolution histories of planetary systems.
   Recent detections of gas in these disks have received considerable attention, 
   as its origin ties up ongoing disk evolution and the present composition of planet-forming materials.}
   {Observations of the CO gas alone, however, cannot reliably differentiate between two leading, competing hypotheses:
   (1) the observed gas is the leftover of protoplanetary disk gas,
   and (2) the gas is the outcome of collisions between icy bodies.
   We propose that such differentiation may become possible by observing cold water vapor.}
   {Order-of-magnitude analyses and comparison with existing observations are performed.}
   {We show that different hypotheses lead to different masses of water vapor.
   This occurs because, for both hypotheses, 
   the presence of cold water vapor is attributed to photodesorption from dust particles by attenuated interstellar UV radiation.
   Cold water vapor cannot be observed by current astronomical facilities as most of its emission lines fall in the far-IR (FIR) range.}
   {This work highlights the need for a future FIR space observatory to reveal the origin of gas in debris disks and the evolution of planet-forming disks in general.}


   \maketitle
%

\section{Introduction}

Debris disks are an integral component of (exo)planetary systems
and are recognized as a good record holder of important insights about their origins and dynamical histories 
\citep[e.g.,][]{2008ARA&A..46..339W,2018ARA&A..56..541H}.
They emerge after most of the gas in protoplanetary disks is gone, that is, systems are older than a few Myrs (typically $10-100$ Myrs). 
These disks are viewed as massive or younger counterparts of the asteroid belt and the Kuiper belt in the present-day solar system.

Debris disks are, by definition, gas-free or at least gas-poor,
and hence detections of gas in the disks have been one fundamental question \citep{1975ApJ...197..137S,1985ApJ...293L..29H}.\footnote{
Historically, the unusual presence of atomic absorption lines in stellar spectra was claimed by \citet{1975ApJ...197..137S}.}
The importance of this problem has recently been boosted up by unexpected detections of CO gas in $\sim$ 20 relatively young ($ \sim10-100$ Myr) debris disks 
\citep[e.g.,][]{2014Sci...343.1490D,2017ApJ...849..123M,2017MNRAS.469..521K,2020MNRAS.492.4409M,2022MNRAS.509..693R}.
The successful detections of gas in debris disks are still very limited as high sensitivity observations with long integration times are required even for known massive debris disks.
The advent of {\it Herschel} and ALMA made it possible to conduct such observations. 
These observations detect the absorption and emission lines originating from atomic and molecular gas in debris disks,
including C, O, Na, Mg, Ca, Mn, Fe, Ni, Ti, Cr \citep[e.g.,][]{1985ApJ...293L..29H,2004A&A...413..681B}, and CO \citep[e.g.,][]{2014Sci...343.1490D,2017ApJ...839...86H}. 
For emission lines, CO gas is the only confirmed molecule so far.  
Ongoing JWST observations aim to discover other molecules, including hot water vapor.

\begin{figure*}
\begin{minipage}{17cm}
\begin{center}
\includegraphics[width=18cm]{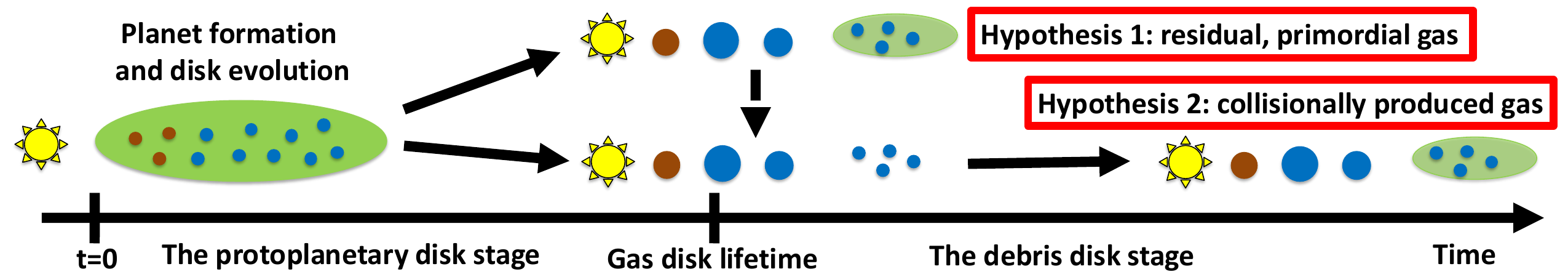}
\caption{A schematic diagram of time-evolution of planet-forming disks.
The host star is denoted by the star symbol, planet-forming materials are represented by the brown and blue dots,
and disk gas by the green ovals.
Hypothesis 1 considers that observed gas is the leftover of primordial gas,
while Hypothesis 2 considers that the gas is produced by collisions among icy bodies.}
\label{fig1}
\end{center}
\end{minipage}
\end{figure*}

Two competing hypotheses are currently proposed to understand the origin of gas in debris disks 
\citep[e.g.,][Figure \ref{fig1}]{2016MNRAS.461..845K,2020MNRAS.492.4409M,2021ApJ...915...90N,2022MNRAS.510.1148S}:
the first one is that observed gas may be leftover from massive gaseous protoplanetary disks, i.e., of primordial origin,
and the competing one is that the gas may be recently generated by colliding planetesimals after gas disks dissipate, that is, it is of secondary origin.\footnote{
For the secondary origin, another hypothesis proposed in the literature is that the observed gas originates 
either from evaporation of cometary bodies due to the host stellar radiation \citep{1990A&A...236..202B} 
or from planetesimals degassing driven by radiogenic heating \citep{2021MNRAS.505.5654D,2023MNRAS.526.3115B}.} 
We hereafter refer to these two hypotheses as Hypothesis 1 and 2, respectively.
A current, outstanding issue is that observations of CO gas alone cannot reliably determine which hypothesis is dominant.

We here propose that water vapor can serve as a better probe to differentiate the origin of gas in debris disks.
By performing order-of-magnitude calculations and comparing them with existing observations,
we show that the mass of cold water vapor as a function of CO mass exhibits a non-monotonic distribution,
leading to better differentiation.
The detection of such water vapor will become possible only by a future far-IR (FIR) space observatory.
This work therefore supports development of a FIR space probe mission to reveal the origin of gas in debris disks and the evolution of planet-forming disks in general.

\section{Determining the origin of gas} \label{sec:mod}

\subsection{Basic picture} \label{sec:pic}

We begin with the introduction of the basic picture of the problem.

The origin of gas in debris disks may be explained by two hypotheses, 
which correspond to two distinct environments achieved during disk evolution (Figure \ref{fig1}): 
Hypothesis 1 focuses on the end stage of protoplanetary disks, where primordial disk gas is still present,
and Hypothesis 2 targets debris disks, where there is no primordial gas, and if gas is present, it is produced by collisions between icy bodies.

In the protoplanetary disk-like environment, the presence of CO gas (and dust) is confirmed by detections of their emission;
most of CO is in the gas-phase when the temperature is higher than its freezeout temperature of $\sim 20$ K.
This indicates that the abundance of observed CO gas is regulated predominantly by gas phase chemistry.
On the other hand, there is no primordial gas in the gas-poor environment like debris disks.
Therefore, gas is produced by collisions among icy bodies.\footnote{
Release mechanisms of CO gas remain uncertain.
In the literature, photodesorption from collisionally produced dust grains \citep{2007A&A...475..755G}, 
vaporization of dust grains due to high velocity collisions \citep{2007ApJ...660.1541C},
and collisions between cometary bodies \citep{2012ApJ...758...77Z} are proposed.}
This suggests that the abundance of observed CO gas is controlled by physical and chemical processes operating in the solid phase;
freezeout of released CO gas is negligible due to its lower freezeout temperature than the surrounding environment.
Thus, difficulty of constraining the origin of gas in debris disks by CO gas alone is attributed to 
the underlying conditions (i.e., gas vs solid phases) considered in the two hypotheses.

Here, we propose to focus on cold water vapor.
In the protoplanetary disk-like environment, the so-called hot/warm and cold water vapor are present;
the former has temperatures of $\sim 200-1500$ K,\footnote{
The presence of hot/warm water is confirmed by recent JWST observations, 
and its presence can be explained by sublimation \citep[e.g.,][]{2023ApJ...957L..22B}.} 
and the latter temperatures of $\lesssim 100$ K.
We currently consider outer parts of disks, where observed CO gas and solids reside, and hence it is reasonable to focus on cold water vapor.
The temperature of cold water vapor is well below the freezeout temperature of water,
and hence its origin is attributed to photodesorption from dust grains due to attenuated interstellar UV radiation \citep{2000ApJ...544..903W,2005ApJ...635L..85D}.
The presence of cold water vapor is already confirmed by observations of protoplanetary disks \citep[e.g.,][]{2011Sci...334..338H}.
Any detection of water vapor in debris disks is not claimed in the literature yet.
If present, it should be classified as cold water vapor due to the location of debris disks from the host star;
water vapor produced by collisions between icy bodies remains in the gas phase because photodesorption prevents complete freezeout.\footnote{
We confirm in Appendix \ref{sec:app_timescale} that 
photodesorption and photodissociation are the dominant processes to determine the abundance of cold water vapor in typical debris disks.}

In summary, focusing on cold water vapor enables unifying the fundamental process operating in the two hypotheses,
and hence the origin of gas in debris disks is better constrained.
In the following, we demonstrate this, using order-of-magnitude analyses.

\subsection{Order-of-magnitude analysis} \label{sec:analysis}

First, it is useful to estimate when the protoplanetary disk stage ends and the debris disk stage starts.
Such a transition occurs when photoevaporation plays the dominant role in dispersal of primordial disk gas \citep[e.g.,][]{1994ApJ...428..654H,2001MNRAS.328..485C,2014prpl.conf..475A}.

Given that a typical photoevaporation rate is $\sim 10^{-10} M_{\odot}$ yr$^{-1}$ and the transition phase may last for $\sim 10^5$ yr 
\citep[e.g.,][]{2014prpl.conf..475A},\footnote{These typical values would be reasonable for disks around T Tauri stars. 
For disks around Herbig Ae/Be stars, however, some deviation would be possible due to inefficient photoelectric heating.
As a result, Hypothesis 1 may be preferred to better reproduce observational trends \citep[e.g.,][]{2023ApJ...959L..28N}.}
a typical value of the gas mass at the transition ($M_{\rm gas, tran}$) is given as 
$M_{\rm gas, tran} \sim  10^{-10} M_{\odot} \mbox{ yr}^{-1} \times 10^5 \mbox{ yr} \sim 1 M_{\oplus}$.
A characteristic value of observed CO-gas mass at the transition ($\overline{M}_{\rm CO, tran}$) is then written as
\begin{equation}
\label{eq:MCO_tran}
\overline{M}_{\rm CO, tran} \simeq  f_{\rm CO, gas} M_{\rm gas, tran} \sim 10^{-3} M_\oplus,
\end{equation}
where
\begin{eqnarray}
\label{eq:f_COgas}
f_{\rm CO, gas}  & \equiv & \frac{M_{\rm CO, gas}}{M_{\rm gas}} \simeq \frac{M_{\rm CO, gas}}{M_{\rm H_2}} 
                               =  \frac{m_{\rm CO} N_{\rm CO}}{m_{\rm H_2} N_{\rm H_2}} \sim 10^{-3} \\ \nonumber
                           & \sim &  \frac{\overline{M}_{\rm CO, tran}}{M_{\rm gas, tran}}, 
\end{eqnarray}
$M_{\rm gas}$, $M_{\rm H_2}$, and $M_{\rm CO, gas}$ are the total, molecular hydrogen, and CO gas masses, respectively,
and $N_{\rm CO}/N_{\rm H_2} \simeq 10^{-4}$ is a typical value inferred for the interstellar medium \citep[ISM, e.g.,][]{2013ARA&A..51..207B}.
Interestingly, this simple estimate is broadly consistent with a detailed estimate derived from disk observations \citep{2023ApJ...951..111C}.

We now compute the mass ratio of CO gas to dust ($M_{\rm solid}$) for both hypotheses 
to examine how CO gas is used to differentiate these hypotheses.
For Hypothesis 1, the mass ratio can be written as
\begin{equation}
\label{eq:CO_Solid_hyp1a}
\left. \frac{\overline{M}_{\rm CO}}{\overline{M}_{\rm solid}} \right|_{\rm Hyp1} \simeq \frac{M_{\rm CO, gas}}{M_{\rm solid}}
                                                                                                        = \frac{M_{\rm CO, gas}}{M_{\rm gas}} \frac{M_{\rm gas}}{M_{\rm solid}} \simeq f_{\rm CO, gas}f_{\rm GtoS} ,
\end{equation}
where $\overline{M}_{\rm CO}$ and $\overline{M}_{\rm solid}$ are the observed CO gas and dust masses, respectively,
and the mass ratio of gas to solids is often called the gas-to-solid (or dust) ratio: 
\begin{equation}
f_{\rm GtoS} \equiv \frac{M_{\rm gas}}{M_{\rm solid}} \sim 10^{2},
\end{equation}
where the typical value (i.e., $f_{\rm GtoS} \sim 10^{2}$) widely used for the initial stage of protoplanetary disks is adopted.
Consequently,
\begin{equation}
\label{eq:CO_Solid_hyp1}
\left. \frac{\overline{M}_{\rm CO}}{\overline{M}_{\rm solid}} \right|_{\rm Hyp1} \simeq 10^{-1} \left( \frac{f_{\rm CO, gas}}{10^{-3}} \right)  \left( \frac{f_{\rm GtoS}}{10^2} \right).
\end{equation}

Our calculations thus suggest that Hypothesis 1 predicts that $\overline{M}_{\rm CO}/\overline{M}_{\rm solid} \sim 0.1$ in the protoplanetary disk-like environment.
Possible variations of $\overline{M}_{\rm CO}/\overline{M}_{\rm solid}$ are discussed in the next section.
It should be noted that in protoplanetary disks, 
solids may be in the form of dust as well as much larger bodies such as planetesimals and (proto)planets.
Accordingly, our assumption (i.e., $\overline{M}_{\rm solid} \simeq M_{\rm solid}$) in equation (\ref{eq:CO_Solid_hyp1a}) leads to an upper limit for the observed mass ratio.

For Hypothesis 2,
we may assume that the observed mass ratio of CO to solids reflects the bulk composition of icy bodies on average
since collisions among various bodies would occur.
Such colliding bodies may have the mass ratio of ice to refractory materials, also known as the ice-to-rock ratio:
\begin{equation}
f_{\rm ItoR} \equiv \frac{M_{Z, \rm ice}}{M_{Z, \rm ref}} \sim 0.1,
\end{equation}
where a characteristic value of $f_{\rm ItoR} (\sim 10^{-1}$) inferred from comets and Kuiper Belt objects (KBOs) is adopted
\citep[e.g.,][]{2019A&A...630A..19B,2019MNRAS.482.3326F,2020Icar..33513421C,2020FrP.....8..227M,2020SSRv..216...44C,2022MNRAS.509.3065D}.
Then $\left. \overline{M}_{\rm CO}/\overline{M}_{\rm solid} \right|_{\rm Hyp2}$ is written as
\begin{equation}
\label{eq:CO_Solid_hyp2_prep}
\left. \frac{\overline{M}_{\rm CO}}{\overline{M}_{\rm solid}} \right|_{\rm Hyp2} \simeq \frac{M_{\rm CO, ice}}{M_{Z, \rm ref}} 
                                                     = \frac{M_{\rm CO, ice}}{M_{Z, \rm ice}} \frac{M_{Z, \rm ice}}{M_{Z, \rm ref}} \simeq f_{\rm CO, ice} f_{\rm ItoR}                                    
\end{equation}
where
\begin{equation}
 f_{\rm CO, ice}  \equiv  \frac{M_{\rm CO, ice}}{M_{Z, \rm ice}} 
                           \simeq  \frac{M_{\rm CO, ice}}{M_{\rm H_2O, ice}} \simeq  \frac{m_{\rm CO} N_{\rm CO, ice}}{m_{\rm H_2O} N_{\rm H_2O, ice}} \sim 10^{-1},
\end{equation}
where it is assumed that $M_{Z, \rm ice} \simeq M_{\rm H_2O, ice}$,
which is reasonable for comets observed in the solar system,
and a characteristic value of $N_{\rm CO, ice} /  N_{\rm H_2O, ice} (\simeq 10^{-1})$ inferred for them is used \citep[e.g.,][]{2011ARA&A..49..471M,2022PSJ.....3..247H}.\footnote{
The lower abundance of CO ice than that of water ice could suggest that the majority of comets in the solar system formed within the CO snow line.
For extrasolar environments, most of debris disks are observed around 100 au away from the host star \citep[e.g.,][]{2018ARA&A..56..541H}.
Under the optically thin assumption, the temperature at the location of debris disks becomes $28$ K around G-type stars, which is higher than the CO freezeout temperature. 
Adopting the abundance ratio inferred from solar system comets would thus be a reasonable first step for extrasolar environments.
On the other hand, if debris disks are located around 200 au, then the temperature goes down to $\sim 20$ K and becomes comparable to the CO freezeout temperature.
For this case, the abundance ratio of CO ice to water ice may become comparable, and differentiation discussed in Section \ref{sec:water} may be less clear.}

Consequently,
\begin{equation}
\label{eq:CO_Solid_hyp2}
\left. \frac{\overline{M}_{\rm CO}}{\overline{M}_{\rm solid}} \right|_{\rm Hyp2}  \simeq 10^{-2} \left( \frac{f_{\rm CO, ice}}{10^{-1}} \right) \left( \frac{f_{\rm ItoR}}{10^{-1}} \right).
\end{equation}
We emphasize that this is a very simple estimate derived from the order-of-magnitude analysis.
In reality, different bodies have different sizes and compositions, resulting in different production rates of the CO gas.
Also, care is needed because values used in equation (\ref{eq:CO_Solid_hyp2}) come mainly from coma gas abundance ratios, which do not necessarily reflect nucleus ice abundance ratios.

Thus, our order-of-magnitude analysis indicates that the two hypotheses lead to different values of the CO-to-solid mass ratio,
and the transition would occur around $M_{\rm CO} \sim 10^{-3} M_{\oplus}$.
It is, however, important to recognize that certain characteristic values are used to obtain the above theoretical estimates.
These values may vary significantly for different targets.

\subsection{Comparison with observations}

Here, we examine how different disk properties would change the above theoretical estimates.
By comparing with existing observational data, 
we explore how reliable the CO-to-solid mass ratio is to differentiate the two hypotheses.

\begin{figure}
\begin{center}
\includegraphics[width=9cm]{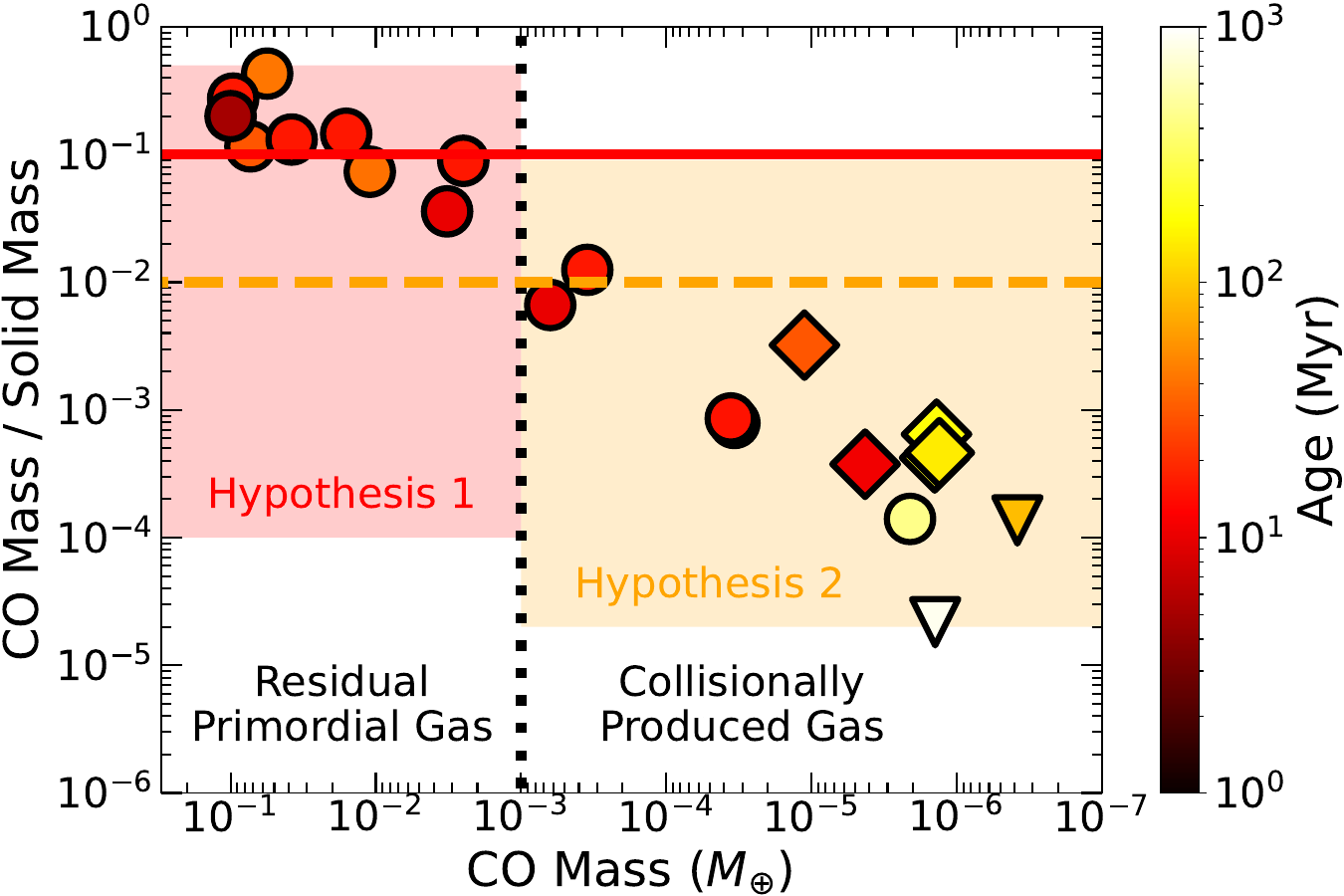}
\caption{The mass ratio of CO gas to solids as a function of CO gas mass.
The theoretical estimates derived from the order-of-magnitude analysis are denoted by the red solid, the orange dashed, and the black dotted lines, respectively 
(see equations (\ref{eq:CO_Solid_hyp1}),  (\ref{eq:CO_Solid_hyp2}), and  (\ref{eq:MCO_tran})).
For comparison, the observed data are included \citep{2017ApJ...849..123M,2019ApJ...884..108M,2020A&A...635A..94D,2022MNRAS.509..693R};
the circles represent targets that exhibit the detections of both CO gas and dust, 
the triangles are for targets that have the detection of only dust (i.e., the upper limit for CO-gas mass),
and the diamonds are for targets with upper limits for both CO gas and dust.
The color of each data point corresponds to the age of the target as shown in the color bar on the right.
Different disk parameters lead to different values of the CO-to-solid mass ratio (Table \ref{table1}).
The red and orange shaded regions represent possible ranges corresponding to the Hypotheses 1 and 2, respectively.
It is clear that the CO-to-solid mass ratio is not a reliable probe to tightly constrain the origin of the observed gas.}
\label{fig2}
\end{center}
\end{figure}

We first compare the theoretical estimates derived in the above section with observational data.
Figure \ref{fig2} depicts three estimates ($\left. \overline{M}_{\rm CO}/  \overline{M}_{\rm solid} \right|_{\rm Hyp1}$, 
$\left.  \overline{M}_{\rm CO}/  \overline{M}_{\rm solid} \right|_{\rm Hyp2}$, and $ \overline{M}_{\rm CO, tran}$)
by the red solid, the orange dashed, and the black dotted lines, respectively (see equations (\ref{eq:CO_Solid_hyp1}),  (\ref{eq:CO_Solid_hyp2}), and  (\ref{eq:MCO_tran})).
The observational data are adopted from \citet{2022MNRAS.509..693R}, wherein the mass estimates of CO and dust are compiled for 21 targets,
using their observations as well as collecting data from the literature \citep[][]{2017ApJ...849..123M,2019ApJ...884..108M,2020A&A...635A..94D}.
As expected, the comparison suggests that Hypothesis 1 works well for CO-gas rich systems, while Hypothesis 2 can explain some of CO-gas poor systems.
It is, however, clear that adopting certain characteristic values cannot reproduce all the observations.

We then explore variation of disk parameters.
Possible ranges of disk parameters ($f_{\rm CO, gas}$, $f_{\rm GtoS}$, $f_{\rm CO, ice}$, and $f_{\rm ItoR} $) are motivated by previous studies (Table \ref{table1}).
One immediately notices that disk parameters span a wide range of values,
and as a result, both hypotheses cover large areas of Figure \ref{fig2};
such areas are denoted by the red and orange shaded regions for Hypotheses 1 and 2, respectively.
This is direct reflection of the wide range of disk parameters.

\begin{table*}
\begin{minipage}{17cm}
\centering
\caption{Possible values of the mass ratios between CO and solids and between H$_2$O and CO}
\label{table1}
{
\begin{tabular}{c|c|c|c}
\hline 
Hypothesis    &  Possible ranges                                                             &  Resulting values                                                                &  Reference       \\ \hline 
Hypothesis 1 & $10^{-4} \lesssim f_{\rm CO, gas} \lesssim 10^{-3}$        & $10^{-4} \lesssim M_{\rm CO}/M_{\rm solid }\lesssim 0.5 $ & (1) \\ \   
                      & $1 \lesssim f_{\rm GtoS} \lesssim 5 \times10^{2}$           & $ 2 \times 10^{-6} \lesssim M_{\rm H_2O}/M_{\rm CO }\lesssim 10^{-2} $ &      \\ \hline    
Hypothesis 2 & $ 4 \times 10^{-3} \lesssim f_{\rm CO, ice} \lesssim 0.3$  & $2 \times 10^{-5} \lesssim M_{\rm CO}/M_{\rm Solid}\lesssim 9 \times 10^{-2} $  & (2) \\
                      & $5 \times 10^{-3} \lesssim f_{\rm ItoR} \lesssim 0.3$         &  $10^{-4} \lesssim M_{\rm H_2O}/M_{\rm CO }\lesssim  0.5 $  &  \\
\hline              
\end{tabular}

(1) The range of $f_{\rm CO, gas}$ comes from \citet{2013ARA&A..51..207B,2013ApJ...776L..38F,2014ApJ...794..160F,2021AJ....161..217C}, 
while that of $f_{\rm GtoS}$ is from \citet{2017A&A...599A.113M}. \\
(2) The range of $f_{\rm CO, ice}$ is motivated by \citet{2011ARA&A..49..471M,2022PSJ.....3..247H}, 
while that of $f_{\rm ItoR} $ is by \citet{2019MNRAS.482.3326F,2020SSRv..216...44C}.
}
\end{minipage}
\end{table*}

The wide range of the CO-to-solid mass ratio indicates two points:
The first one is that even if different targets have different disk properties,
high CO masses tend to be explained well by Hypothesis 1, that is, observed gas is very likely of primordial origin.
On the other hand, low CO masses are better explained by Hypothesis 2, that is, the gas is likely produced by collisions among icy bodies.
This thus suggests that the order-of-magnitude analysis is useful for developing a basic understanding.
The second point is, however, that both Hypotheses 1 and 2 can cover the same values of the CO-to-solid mass ratio (i.e., from $10^{-4}$ to 0.1),
and hence it is not straightforward to reliably differentiate the origin of observed gas.
Distinction between Hypotheses 1 and 2 can be done by specifying $ \overline{M}_{\rm CO, tran}$ (equation (\ref{eq:MCO_tran})),
yet the monotonic decrease of the CO-to-solid mass ratio with the CO mass makes the transition ambiguous.
If non-monotonic distributions would be predicted by another probe, more reliable determination would be possible.

Thus, our analysis shows that both the hypotheses can reproduce existing observational data,
and hence they are both viable.
However, the resulting mass ratios indicate that the origin of gas in debris disks cannot be determined reliably from the value of $\overline{M}_{\rm CO}/\overline{M}_{\rm solid}$.

\subsection{The importance of water vapor} \label{sec:water}

Here, we discuss how the degeneracy discussed above can be broken and the origin of observed gas can be constrained better.
In this work, we propose to focus on water vapor, resulting in a non-monotonic behavior of the mass ratio.

As discussed in Section \ref{sec:pic}, 
the presence of water vapor can be expected under both hypotheses and explained by photodesorption, which works as follows:
at the disk surface, unattenuated UV radiation photodissociates water gas entirely,
that is, no water vapor is present.
On the other hand, UV radiation is completely attenuated at the disk midplane, and water gas completely freezes onto dust grains.
In between, attenuated UV flux is too weak to photodissociate water gas entirely, 
but is still strong enough for photodesorption to prevent complete freezeout of water molecules onto dust grains \citep{2009ApJ...690.1497H}.
In such photodesorption-dominated regions, cold water vapor can be present.

The abundance of cold water vapor is then essentially insensitive to the UV flux \citep{2005ApJ...635L..85D,2014ApJ...792....2D}.
This is because the abundance is determined mainly by photodesorption and photodissociation, 
which are both controlled by the UV flux.
Consequently, the abundance of water vapor ($N_{\rm H_2O,vap}$) is computed as
\begin{equation}
\label{eq:balance}
\sigma_{\rm photo} N_{\rm H_2O,vap}  = \sigma_{\rm dust} Y  N_{\rm dust} , 
\end{equation}
where $\sigma_{\rm photo}$ is the photodissociation cross-section of water, 
$ \sigma_{\rm dust}=\pi r_{\rm dust}^2$ is the cross-section of dust particles with radius of $r_{\rm dust}$,
$Y$ is the photodesorption yield of water, 
and $N_{\rm dust}$ is the abundance of dust particles contributing to photodesorption,
which may differ for different hypotheses.
In the following calculations, we adopt the standard values of $Y$ \citep[$\sim 10^{-3}$,][]{1995Natur.373..405W,2009ApJ...693.1209O} 
and of $\sigma_{\rm photo} (= R_0/F_0 \sim 10^{-17}$ cm$^{2}$), 
where $R_0 \sim 10^{-9}$ s$^{-1}$ is the unshielded photodissociation rate of water at $G_0=1$\footnote{The UV flux is often normalized in units of $G_0$, 
where 1 $G_{0} =1.6 \times 10^{-3}$ erg cm$^{-2}$ s$^{-1}$ is the flux integrated over the range from 6 to 13.6 eV \citep{1968BAN....19..421H}.} \citep[e.g.,][]{2009ApJ...690.1497H}, 
and $F_0 \sim 10^8$ photons cm$^{-2}$ s$^{-1}$ is the flux of UV photons at $G_0=1$.

It should be noted that the presence of CO gas is confirmed only for debris disks, 
where UV radiation from the host star is weak enough that interstellar UV radiation plays a major role in photodissociation \citep{2017MNRAS.469..521K}.
Hence, we focus on interstellar UV radiation in this work.
We confirm that our abundance estimates do not change very much even if stellar UV radiation plays a dominant role (Appendix \ref{sec:app_UV}).

We now compute the mass ratio of water vapor ($\overline{M}_{\rm H_2O}$) to CO gas.
For hypothesis 1, $N_{\rm dust}$ in equation (\ref{eq:balance}) can be expressed as $N_{\rm dust}  \equiv  f_{\rm dust} M_{\rm solid}/ m_{\rm dust}$,
where $f_{\rm dust}$ is a factor taking account of the dust abundance relative to the total solid abundance,
$m_{\rm dust}= 4 \pi \rho_{\rm dust} r_{\rm dust}^3/3$ is the mass of dust particles that have the bulk density of $\rho_{\rm dust}$.
Then, the ratio of $N_{\rm H_2O,vap}$ to $ N_{\rm H_2}$ can be written as (equation (\ref{eq:balance}))
\begin{equation}
\label{eq:M_H2O_CO_hyp1}
\frac{ N_{\rm H_2O,vap} }{ N_{\rm H_2}} = Y \frac{\sigma_{\rm dust}}{\sigma_{\rm photo}} \frac{N_{\rm dust}}{N_{\rm H_2}} 
                      \sim 10^{-9} \left( \frac{f_{\rm GtoS}}{10^2} \right)^{-1} \left( \frac{r_{\rm dust}}{0.1~ \mu\mbox{m}} \right)^{-1} ,
\end{equation}
where it is assumed that dust particles with a characteristic size of $r_{\rm dust} (\sim 0.1 ~ \mu$m) are most important for photodesorption.
In order to better reproduce the result of \citet[see their equation (29)]{2014ApJ...792....2D} and the observation of cold water vapor in a protoplanetary disk \citep{2011Sci...334..338H},
the abundance of such dust particles is set at 1 \% (i.e., $f_{\rm dust}=10^{-2}$), 
which reflects the effect of dust growth and settling.

Consequently, the ratio ($\overline{M}_{\rm H_2O}/\overline{M}_{\rm CO}$) for Hypothesis 1 can be written as
\begin{eqnarray}
\label{eq:M_H2O_CO_hyp1}
\left. \frac{\overline{M}_{\rm H_2O}}{\overline{M}_{\rm CO}} \right|_{\rm Hyp1} 
            & \simeq & \frac{m_{\rm H_2O} N_{\rm H_2O,vap}}{M_{\rm CO,gas}}  
                           =  \frac{M_{\rm H_2}}{M_{\rm CO,gas}} \frac{ m_{\rm H_2O} N_{\rm H_2O,vap}}{m_{\rm H_2} N_{\rm H_2}} \\ \nonumber
              & \sim  & 10^{-5}  \left( \frac{f_{\rm CO, gas} }{10^{-3}} \right)^{-1}  \left( \frac{f_{\rm GtoS}}{10^2} \right)^{-1} \left( \frac{r_{\rm dust}}{0.1 \mu\mbox{m}} \right)^{-1} .
\end{eqnarray}

For Hypothesis 2, both gas and dust particles originate from collisions between icy bodies.
Accordingly, $N_{\rm dust}$ is given as $N_{\rm dust}  \equiv  f_{\rm dust} M_{\rm Z, ref}/m_{\rm dust}$.
Then, the ratio of $N_{\rm H_2O,vap}$ to $ N_{\rm H_2O, ice}$ can be written as (equation (\ref{eq:balance}))
\begin{equation}
\label{eq:N_H2Ovap_hyp2}
\frac{ N_{\rm H_2O,vap} }{ N_{\rm H_2O, ice}}  = Y \frac{\sigma_{\rm dust}}{\sigma_{\rm photo}} \frac{N_{\rm dust}}{N_{\rm H_2O, ice}} 
                                                                            \sim 10^{-4} \left( \frac{f_{\rm ItoR}}{10^{-1}} \right)^{-1} \left( \frac{r_{\rm dust}}{1~ \mu\mbox{m}} \right)^{-1} ,
\end{equation}
where it is assumed that $N_{\rm H_2O, ice} \simeq M_{Z,\rm ice}/m_{\rm H_2O}$ as inferred for comets
and that dust particles with a typical size of $r_{\rm dust} (\sim 1 ~ \mu$m) contribute most to photodesorption;
their size is constrained by the blowout size below which dust particles are expelled directly by stellar radiation pressure forces.
The abundance of such particles is set at unity here (i.e., $f_{\rm dust}=1$), 
following the consideration adopted when computing equation (\ref{eq:CO_Solid_hyp2_prep});
the observed mass ratio should reflect the bulk composition of icy bodies on average since collisions among various bodies would occur.
A low value of $N_{\rm H_2O,vap} / N_{\rm H_2O, ice}$ implies that most of the water would be present at the midplane region,
where water vapor freezeouts onto existing dust grains or used as nucleation of new icy grains.

As a result, the ratio ($\overline{M}_{\rm H_2O}/\overline{M}_{\rm CO}$) for Hypothesis 2 can be written as
\begin{eqnarray}
\label{eq:M_H2O_CO_hyp2}
\left. \frac{\overline{M}_{\rm H_2O}}{\overline{M}_{\rm CO}} \right|_{\rm Hyp2} 
                               & \simeq & \frac{m_{\rm H_2O} N_{\rm H_2O,vap}}{M_{\rm CO, ice}} 
                               =  \frac{M_{\rm H_2O, ice}}{M_{\rm CO,ice}} \frac{ N_{\rm H_2O,vap}}{ N_{\rm H_2O,ice}} \\ \nonumber
                               & \sim &10^{-3}  \left( \frac{f_{\rm CO, ice} }{10^{-1}} \right)^{-1} \left( \frac{f_{\rm ItoR}}{10^{-1}} \right)^{-1} \left( \frac{r_{\rm dust}}{1~ \mu\mbox{m}} \right)^{-1}.
\end{eqnarray}

\begin{figure*}
\begin{minipage}{17cm}
\begin{center}
\includegraphics[width=8cm]{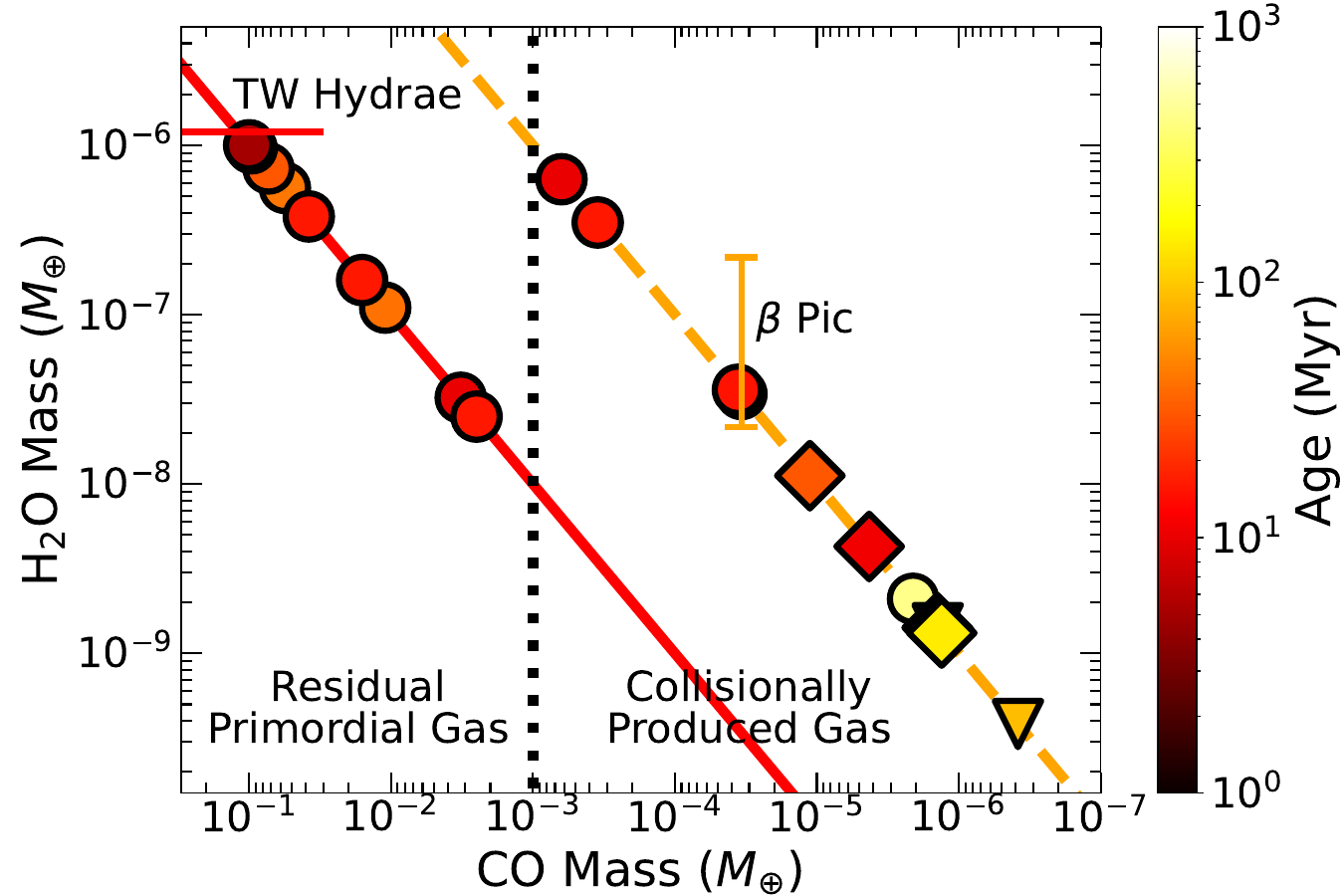}
\includegraphics[width=8cm]{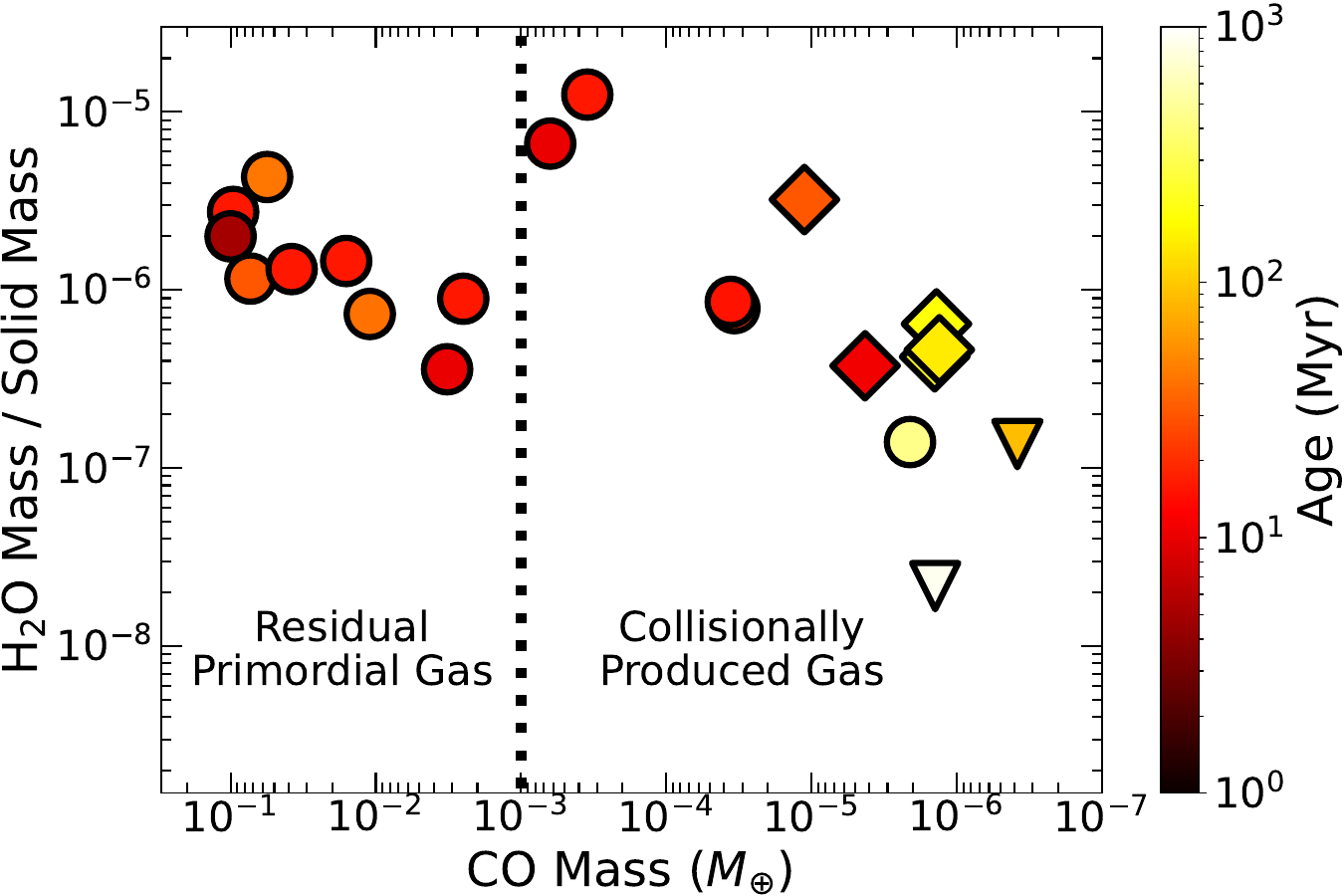}
\caption{The predicted water vapor mass and its mass ratio relative to the solid mass as a function of CO mass in the left and right panels, respectively.
All the symbols are defined in Figure \ref{fig2}.
As an example, the H$_2$O-to-CO conversion factor is switched 
from equation (\ref{eq:M_H2O_CO_hyp1}) for Hypothesis 1 to equation (\ref{eq:M_H2O_CO_hyp2}) for Hypothesis 2 at the transition CO mass of $10^{-3}$;
for clear comparison, these conversion factors are shown by the red solid and the orange dashed lines for Hypotheses 1 and 2, respectively (left panel).
Such switching results in a clear jump in the H$_2$O mass and its mass ratio.
On the left panel, tentative constrains derived from previous observations are also plotted;
the water vapor mass estimated for the protoplanetary disk around TW Hydrae is denoted by the horizontal, red solid line segment,
while that for the debris disk around $\beta$ Pic is represented by the error bar.
The non-monotonic distributions in the water vapor mass and its mass ratio relative to the solid mass enable better determination of the origin of gas in debris disks.}
\label{fig3}
\end{center}
\end{minipage}
\end{figure*}

We finally compute the predicted H$_2$O mass as a function of CO mass and its mass ratio relative to the solid mass.
Figure \ref{fig3} shows the results.
As an example, 
the H$_2$O-to-CO conversion factor (equation (\ref{eq:M_H2O_CO_hyp1}) vs equation (\ref{eq:M_H2O_CO_hyp2})) 
is switched at $\overline{M}_{\rm CO} \sim 10^{-3} M_{\oplus}$ (equation (\ref{eq:MCO_tran})).
It is clear that such switching leads to a noticeable jump in $\overline{M}_{\rm H_2O} $ and $\overline{M}_{\rm H_2O} /\overline{M}_{\rm solid}$ at that location.
The resulting non-monotonic distribution differs from the monotonic distribution of $\overline{M}_{\rm CO} /\overline{M}_{\rm solid}$ (c.f., Figure \ref{fig2})
and statistically enables better specification of the value of  $\overline{M}_{\rm CO} $, 
at which the gas observed in debris disks would be likely of primordial origin or of secondary origin.

Our predications can be tested tentatively by previous observations.
As described below, emission lines from cold water vapor fall in FIR wavelength range,
and {\it Herschel} observations provide some constraints.
For instance, cold water vapor was detected in the protoplanetary disk around TW Hydrae,
and its mass was estimated  to be $1.2 \times 10^{-6} M_{\oplus}$ \citep{2011Sci...334..338H}.
As another example, an upper limit of the water vapor mass is estimated to be $1.2 \times 10^{-8}-1.8 \times 10^{-7} M_{\oplus}$ 
from non-detection observations taken toward the debris disk around $\beta$ pic \citep{2019A&A...628A.127C}.
Importantly, these values are consistent with our estimate (Figure \ref{fig3});
while a number of estimates for the CO gas mass are available for the disk around TW Hydrae,
they have large uncertainties \citep[e.g.,][]{2022ApJ...926L...2T}.
Hence, its location in the $x-$axis is not tightly constrained currently.

The above tentative test is promising.
However, it is natural to anticipate that the two hypotheses would end up with the same value of $\overline{M}_{\rm H_2O} /\overline{M}_{\rm CO}$ for a wide range of disk parameters.
This is confirmed in Table \ref{table1}.
Nonetheless, it is important to recognize that the range of overlapping values shrinks considerably (i.e., from $10^{-4}$ to $10^{-2}$),
compared with the case of $\overline{M}_{\rm CO} /\overline{M}_{\rm solid}$ (i.e., from $10^{-4}$ to $10^{-1}$).
Such a difference arises because the abundance of cold water vapor is determined by photodesorption for both hypotheses,
so that it becomes more sensitive to its origin.
Given the wide range of disk parameters, statistically identifying a non-monotonic behavior would be a key.
In Appendix \ref{sec:app_mc}, we perform simple Monte-Carlo calculations and confirm that such statistical identification would be possible 
if a couple of dozens of targets are observed with the CO and water gas mass sensitivities of $10^{-4} M_{\oplus}$ and of $10^{-9}-10^{-8} M_{\oplus}$ or lower, respectively,
and if these targets have a wide coverage of the CO gas mass.

Thus, our calculations show that water vapor (in combination with CO gas) becomes a better tracer of the origin of gas in debris disks.

\section{Summary and Outlook} \label{sec:disc}

Revealing the origin of gas in debris disks is one outstanding issue in understanding the time-evolution of planet-forming disks.
In the literature, two competing hypotheses are proposed (i.e., the primordial origin vs the secondary origin, Figure \ref{fig1}).
Current observations of CO gas alone cannot distinguish between these hypotheses reliably (Figure \ref{fig2}).
We have demonstrated through order-of-magnitude analyses that cold water vapor can play such a role.
This becomes possible because the abundance of water vapor differs between these two hypotheses considerably 
(equation (\ref{eq:M_H2O_CO_hyp1}) vs equation (\ref{eq:M_H2O_CO_hyp2})).
The resulting distribution of water mass as a function of CO mass shows a non-monotonic behavior (Figure \ref{fig3}),
which can be used to better specify when the relative importance of these hypotheses would switch.
This work therefore demonstrates that the detection of cold water vapor is a key to shedding light on the origin of gas in debris disks.

\begin{figure}
\begin{center}
\includegraphics[width=9cm]{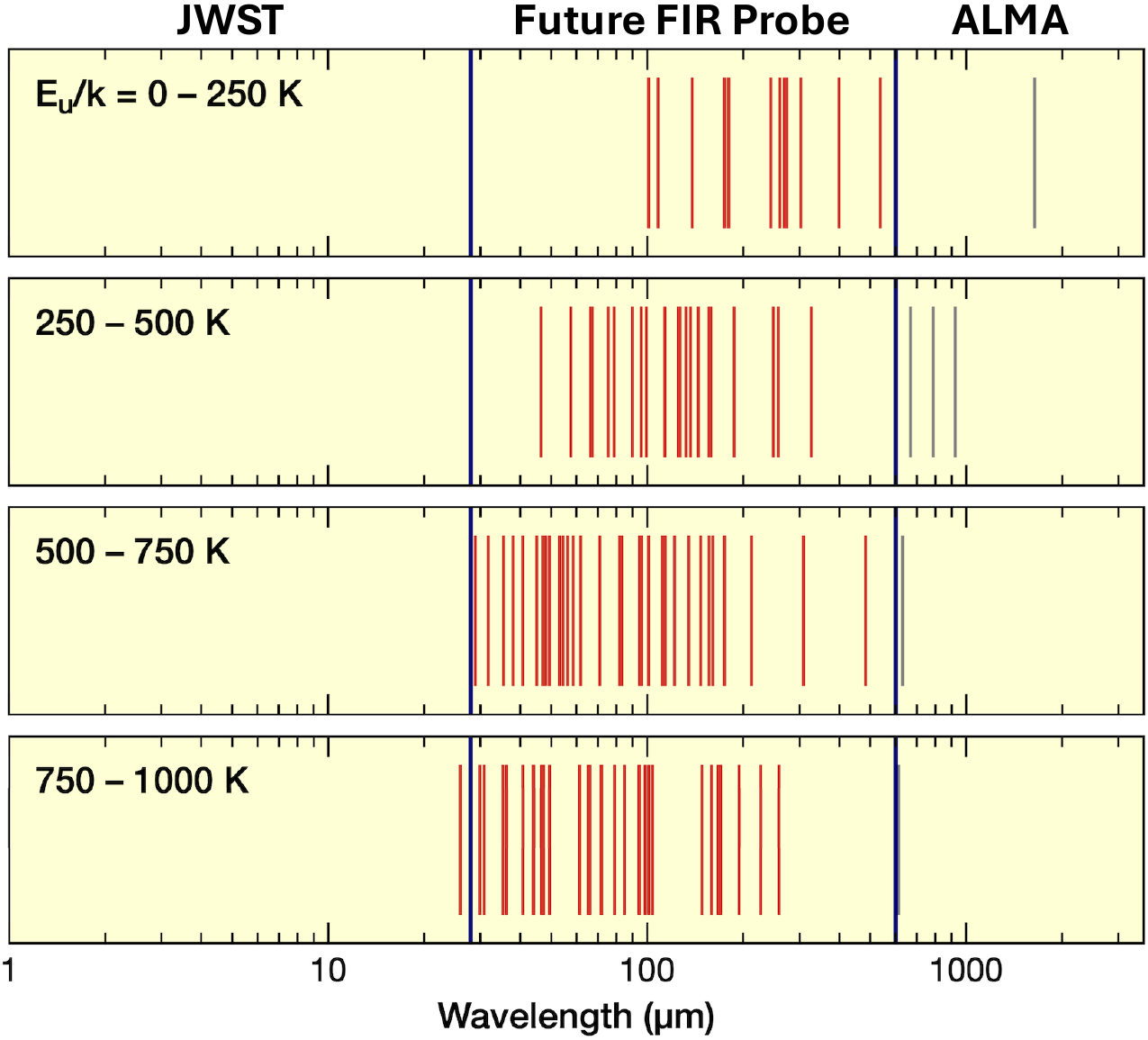}
\caption{Emission lines from water vapor as a function of wavelength.
The excitation temperature increases from the top to the bottom panels.
Different observatories cover different wavelengths;
this plot confirms that water lines detected by JWST come from high excitation temperatures ($\gtrsim 1000$ K, i.e., hot/warm water).
A future FIR space observatory is the most sensitive to cold water vapor.
A spectral resolution of $R \sim$ a few thousands appears to be sufficient for detections.}
\label{fig4}
\end{center}
\end{figure}

Observations of cold water vapor are not possible with currently available telescopes, however.
To illustrate this, Figure \ref{fig4} shows emission lines from water vapor as a function of wavelength.
It is clear that most lines originating from low excitation temperatures fall in the FIR range.
Such a wavelength range may become accessible in the future, thanks to the recent astrophysics probe call by NASA.
This work strongly supports development of a future FIR space observatory.

Historically, such wavelengths were covered by {\it Herschel}.
While a number of debris disks were observed by the space telescope \citep[e.g.,][]{2013A&A...555A..11E,2013MNRAS.428.1263B},
no claim was made in literature that cold water vapor was discovered in these disks.
This is probably because of the sensitivity issue;
as described above, {\it Herschel} detected water vapor in the protoplanetary disk around TW Hydrae.
A similar value of the water vapor mass would be possible for CO-gas rich debris disks (Figure \ref{fig3}).
Non detections might therefore be explained by the fact that while TW Hydrae is about 60 pc away,
these gas-rich debris disks are $>100$ pc away, resulting in decrease of the observable flux by a factor of $4$ or more due to beam dilution.
In order to reliably detect cold water vapor in debris disks and carefully examine the origin,
a desired sensitivity improvement factor over {\it Herschel} would be 100 or more for a future FIR instrument.

We have chosen to adopt the simplified approach in this work as the detection of water vapor in debris disks remains to be performed.
Our assumption likely holds for typical debris disks (Appendix \ref{sec:app_timescale}) and our results are promising for future observations.
However, more detailed work is required to confirm the finding.
It is critical to identify how much gas is released at what size of solids;
if debris disks are massive enough (i.e., $M_{Z, \rm ref} > 0.1 M_{\oplus}$, see Appendix \ref{sec:app_timescale}), 
then different production and destruction processes may become dominated,
so that the resulting water vapor abundance could be very low.
In addition, simultaneous calculations of disk evolution and chemistry are key for Hypothesis 1,
while coupling of gas production calculations by collisions with disk chemistry is crucial for Hypothesis 2.
We have shown that our abundance estimates would be valid for both interstellar and stellar UV radiation (Appendix \ref{sec:app_UV}).
However, the lifetime of water vapor depends on the UV source;
if stellar UV radiation dominates, the lifetime could be very short, and the detection of water vapor may be challenging.
For this case, observations of OH, which is the outcome of photodissociation of water, 
may be used to infer the abundance of water vapor \citep{2021A&A...650A.192T,2024NatAs...8..577Z}.
Clearly, more dedicated investigations remain for future work.

In conclusion, modeling and detections of various molecules shed light on the origin of gas in debris disks,
and this work puts the particular focus on cold water vapor.

\begin{acknowledgements}
The authors thank an anonymous referee for the useful comments on our manuscript.
This research was carried out at the Jet Propulsion Laboratory, California Institute of Technology, 
under a contract with the National Aeronautics and Space Administration (80NM0018D0004) 
and funded through the internal Research and Technology Development program 
as well as the NASA Exoplanets Research Program through grant 23-XRP23\_2-0128.
Y.H. is supported by JPL/Caltech.
D.C.L. acknowledges financial support from the National Aeronautics and Space Administration (NASA) Astrophysics Data Analysis Program (ADAP) .

\end{acknowledgements}

\bibliographystyle{aa}
\bibliography{adsbibliography}

\begin{thebibliography}{57}
\expandafter\ifx\csname natexlab\endcsname\relax\def\natexlab#1{#1}\fi

\bibitem[{{Alexander} {et~al.}(2014){Alexander}, {Pascucci}, {Andrews},
  {Armitage}, \& {Cieza}}]{2014prpl.conf..475A}
{Alexander}, R., {Pascucci}, I., {Andrews}, S., {Armitage}, P., \& {Cieza}, L.
  2014, in Protostars and Planets VI, ed. H.~{Beuther}, R.~S. {Klessen}, C.~P.
  {Dullemond}, \& T.~{Henning}, 475--496

\bibitem[{{Banzatti} {et~al.}(2023){Banzatti}, {Pontoppidan}, {Carr},
  {Jellison}, {Pascucci}, {Najita}, {Mu{\~n}oz-Romero}, {{\"O}berg}, {Kalyaan},
  {Pinilla}, {Krijt}, {Long}, {Lambrechts}, {Rosotti}, {Herczeg}, {Salyk},
  {Zhang}, {Bergin}, {Ballering}, {Meyer}, {Bruderer}, \& {Jdiscs
  Collaboration}}]{2023ApJ...957L..22B}
{Banzatti}, A., {Pontoppidan}, K.~M., {Carr}, J.~S., {et~al.} 2023, \apjl, 957,
  L22

\bibitem[{{Bergin} {et~al.}(2003){Bergin}, {Calvet}, {D'Alessio}, \&
  {Herczeg}}]{2003ApJ...591L.159B}
{Bergin}, E., {Calvet}, N., {D'Alessio}, P., \& {Herczeg}, G.~J. 2003, \apjl,
  591, L159

\bibitem[{{Beust} {et~al.}(1990){Beust}, {Lagrange-Henri}, {Vidal-Madjar}, \&
  {Ferlet}}]{1990A&A...236..202B}
{Beust}, H., {Lagrange-Henri}, A.~M., {Vidal-Madjar}, A., \& {Ferlet}, R. 1990,
  \aap, 236, 202

\bibitem[{{Biver} {et~al.}(2019){Biver}, {Bockel{\'e}e-Morvan}, {Hofstadter},
  {Lellouch}, {Choukroun}, {Gulkis}, {Crovisier}, {Schloerb}, {Rezac}, {von
  Allmen}, {Lee}, {Leyrat}, {Ip}, {Hartogh}, {Encrenaz}, {Beaudin}, \& {MIRO
  Team}}]{2019A&A...630A..19B}
{Biver}, N., {Bockel{\'e}e-Morvan}, D., {Hofstadter}, M., {et~al.} 2019, \aap,
  630, A19

\bibitem[{{Bolatto} {et~al.}(2013){Bolatto}, {Wolfire}, \&
  {Leroy}}]{2013ARA&A..51..207B}
{Bolatto}, A.~D., {Wolfire}, M., \& {Leroy}, A.~K. 2013, \araa, 51, 207

\bibitem[{{Bonsor} {et~al.}(2023){Bonsor}, {Wyatt}, {Marino}, {Davidsson},
  {Kral}, \& {Thebault}}]{2023MNRAS.526.3115B}
{Bonsor}, A., {Wyatt}, M.~C., {Marino}, S., {et~al.} 2023, \mnras, 526, 3115

\bibitem[{{Booth} {et~al.}(2013){Booth}, {Kennedy}, {Sibthorpe}, {Matthews},
  {Wyatt}, {Duch{\^e}ne}, {Kavelaars}, {Rodriguez}, {Greaves}, {Koning},
  {Vican}, {Rieke}, {Su}, {Moro-Mart{\'\i}n}, \& {Kalas}}]{2013MNRAS.428.1263B}
{Booth}, M., {Kennedy}, G., {Sibthorpe}, B., {et~al.} 2013, \mnras, 428, 1263

\bibitem[{{Brandeker} {et~al.}(2004){Brandeker}, {Liseau}, {Olofsson}, \&
  {Fridlund}}]{2004A&A...413..681B}
{Brandeker}, A., {Liseau}, R., {Olofsson}, G., \& {Fridlund}, M. 2004, \aap,
  413, 681

\bibitem[{{Cataldi} {et~al.}(2023){Cataldi}, {Aikawa}, {Iwasaki}, {Marino},
  {Brandeker}, {Hales}, {Henning}, {Higuchi}, {Hughes}, {Janson}, {Kral},
  {Matr{\`a}}, {Mo{\'o}r}, {Olofsson}, {Redfield}, \&
  {Roberge}}]{2023ApJ...951..111C}
{Cataldi}, G., {Aikawa}, Y., {Iwasaki}, K., {et~al.} 2023, \apj, 951, 111

\bibitem[{{Cauley} {et~al.}(2021){Cauley}, {France}, {Herzceg}, \&
  {Johns-Krull}}]{2021AJ....161..217C}
{Cauley}, P.~W., {France}, K., {Herzceg}, G.~J., \& {Johns-Krull}, C.~M. 2021,
  \aj, 161, 217

\bibitem[{{Cavallius} {et~al.}(2019){Cavallius}, {Cataldi}, {Brandeker},
  {Olofsson}, {Larsson}, \& {Liseau}}]{2019A&A...628A.127C}
{Cavallius}, M., {Cataldi}, G., {Brandeker}, A., {et~al.} 2019, \aap, 628, A127

\bibitem[{{Choukroun} {et~al.}(2020){Choukroun}, {Altwegg}, {K{\"u}hrt},
  {Biver}, {Bockel{\'e}e-Morvan}, {Dr{\k{a}}{\.z}kowska}, {H{\'e}rique},
  {Hilchenbach}, {Marschall}, {P{\"a}tzold}, {Taylor}, \&
  {Thomas}}]{2020SSRv..216...44C}
{Choukroun}, M., {Altwegg}, K., {K{\"u}hrt}, E., {et~al.} 2020, \ssr, 216, 44

\bibitem[{{Clarke} {et~al.}(2001){Clarke}, {Gendrin}, \&
  {Sotomayor}}]{2001MNRAS.328..485C}
{Clarke}, C.~J., {Gendrin}, A., \& {Sotomayor}, M. 2001, \mnras, 328, 485

\bibitem[{{Combi} {et~al.}(2020){Combi}, {Shou}, {Fougere}, {Tenishev},
  {Altwegg}, {Rubin}, {Bockel{\'e}e-Morvan}, {Capaccioni}, {Cheng}, {Fink},
  {Gombosi}, {Hansen}, {Huang}, {Marshall}, \& {Toth}}]{2020Icar..33513421C}
{Combi}, M., {Shou}, Y., {Fougere}, N., {et~al.} 2020, \icarus, 335, 113421

\bibitem[{{Czechowski} \& {Mann}(2007)}]{2007ApJ...660.1541C}
{Czechowski}, A. \& {Mann}, I. 2007, \apj, 660, 1541

\bibitem[{{Davidsson}(2021)}]{2021MNRAS.505.5654D}
{Davidsson}, B. J.~R. 2021, \mnras, 505, 5654

\bibitem[{{Davidsson} {et~al.}(2022){Davidsson}, {Samarasinha}, {Farnocchia},
  \& {Guti{\'e}rrez}}]{2022MNRAS.509.3065D}
{Davidsson}, B. J.~R., {Samarasinha}, N.~H., {Farnocchia}, D., \&
  {Guti{\'e}rrez}, P.~J. 2022, \mnras, 509, 3065

\bibitem[{{Dent} {et~al.}(2014){Dent}, {Wyatt}, {Roberge}, {Augereau},
  {Casassus}, {Corder}, {Greaves}, {de Gregorio-Monsalvo}, {Hales}, {Jackson},
  {Hughes}, {Lagrange}, {Matthews}, \& {Wilner}}]{2014Sci...343.1490D}
{Dent}, W.~R.~F., {Wyatt}, M.~C., {Roberge}, A., {et~al.} 2014, Science, 343,
  1490

\bibitem[{{Di Folco} {et~al.}(2020){Di Folco}, {P{\'e}ricaud}, {Dutrey},
  {Augereau}, {Chapillon}, {Guilloteau}, {Pi{\'e}tu}, \&
  {Boccaletti}}]{2020A&A...635A..94D}
{Di Folco}, E., {P{\'e}ricaud}, J., {Dutrey}, A., {et~al.} 2020, \aap, 635, A94

\bibitem[{{Dominik} {et~al.}(2005){Dominik}, {Ceccarelli}, {Hollenbach}, \&
  {Kaufman}}]{2005ApJ...635L..85D}
{Dominik}, C., {Ceccarelli}, C., {Hollenbach}, D., \& {Kaufman}, M. 2005,
  \apjl, 635, L85

\bibitem[{{Du} \& {Bergin}(2014)}]{2014ApJ...792....2D}
{Du}, F. \& {Bergin}, E.~A. 2014, \apj, 792, 2

\bibitem[{{Eiroa} {et~al.}(2013){Eiroa}, {Marshall}, {Mora}, {Montesinos},
  {Absil}, {Augereau}, {Bayo}, {Bryden}, {Danchi}, {del Burgo}, {Ertel},
  {Fridlund}, {Heras}, {Krivov}, {Launhardt}, {Liseau}, {L{\"o}hne},
  {Maldonado}, {Pilbratt}, {Roberge}, {Rodmann}, {Sanz-Forcada}, {Solano},
  {Stapelfeldt}, {Th{\'e}bault}, {Wolf}, {Ardila}, {Ar{\'e}valo}, {Beichmann},
  {Faramaz}, {Gonz{\'a}lez-Garc{\'\i}a}, {Guti{\'e}rrez}, {Lebreton},
  {Mart{\'\i}nez-Arn{\'a}iz}, {Meeus}, {Montes}, {Olofsson}, {Su}, {White},
  {Barrado}, {Fukagawa}, {Gr{\"u}n}, {Kamp}, {Lorente}, {Morbidelli},
  {M{\"u}ller}, {Mutschke}, {Nakagawa}, {Ribas}, \&
  {Walker}}]{2013A&A...555A..11E}
{Eiroa}, C., {Marshall}, J.~P., {Mora}, A., {et~al.} 2013, \aap, 555, A11

\bibitem[{{Favre} {et~al.}(2013){Favre}, {Cleeves}, {Bergin}, {Qi}, \&
  {Blake}}]{2013ApJ...776L..38F}
{Favre}, C., {Cleeves}, L.~I., {Bergin}, E.~A., {Qi}, C., \& {Blake}, G.~A.
  2013, \apjl, 776, L38

\bibitem[{{France} {et~al.}(2014){France}, {Herczeg}, {McJunkin}, \&
  {Penton}}]{2014ApJ...794..160F}
{France}, K., {Herczeg}, G.~J., {McJunkin}, M., \& {Penton}, S.~V. 2014, \apj,
  794, 160

\bibitem[{{Fulle} {et~al.}(2019){Fulle}, {Blum}, {Green}, {Gundlach},
  {Herique}, {Moreno}, {Mottola}, {Rotundi}, \&
  {Snodgrass}}]{2019MNRAS.482.3326F}
{Fulle}, M., {Blum}, J., {Green}, S.~F., {et~al.} 2019, \mnras, 482, 3326

\bibitem[{{Grigorieva} {et~al.}(2007){Grigorieva}, {Th{\'e}bault},
  {Artymowicz}, \& {Brandeker}}]{2007A&A...475..755G}
{Grigorieva}, A., {Th{\'e}bault}, P., {Artymowicz}, P., \& {Brandeker}, A.
  2007, \aap, 475, 755

\bibitem[{{Habing}(1968)}]{1968BAN....19..421H}
{Habing}, H.~J. 1968, \bain, 19, 421

\bibitem[{{Harrington Pinto} {et~al.}(2022){Harrington Pinto}, {Womack},
  {Fernandez}, \& {Bauer}}]{2022PSJ.....3..247H}
{Harrington Pinto}, O., {Womack}, M., {Fernandez}, Y., \& {Bauer}, J. 2022,
  \psj, 3, 247

\bibitem[{{Hobbs} {et~al.}(1985){Hobbs}, {Vidal-Madjar}, {Ferlet}, {Albert}, \&
  {Gry}}]{1985ApJ...293L..29H}
{Hobbs}, L.~M., {Vidal-Madjar}, A., {Ferlet}, R., {Albert}, C.~E., \& {Gry}, C.
  1985, \apjl, 293, L29

\bibitem[{{Hogerheijde} {et~al.}(2011){Hogerheijde}, {Bergin}, {Brinch},
  {Cleeves}, {Fogel}, {Blake}, {Dominik}, {Lis}, {Melnick}, {Neufeld},
  {Pani{\'c}}, {Pearson}, {Kristensen}, {Y{\i}ld{\i}z}, \& {van
  Dishoeck}}]{2011Sci...334..338H}
{Hogerheijde}, M.~R., {Bergin}, E.~A., {Brinch}, C., {et~al.} 2011, Science,
  334, 338

\bibitem[{{Hollenbach} {et~al.}(1994){Hollenbach}, {Johnstone}, {Lizano}, \&
  {Shu}}]{1994ApJ...428..654H}
{Hollenbach}, D., {Johnstone}, D., {Lizano}, S., \& {Shu}, F. 1994, \apj, 428,
  654

\bibitem[{{Hollenbach} {et~al.}(2009){Hollenbach}, {Kaufman}, {Bergin}, \&
  {Melnick}}]{2009ApJ...690.1497H}
{Hollenbach}, D., {Kaufman}, M.~J., {Bergin}, E.~A., \& {Melnick}, G.~J. 2009,
  \apj, 690, 1497

\bibitem[{{Hughes} {et~al.}(2018){Hughes}, {Duch{\^e}ne}, \&
  {Matthews}}]{2018ARA&A..56..541H}
{Hughes}, A.~M., {Duch{\^e}ne}, G., \& {Matthews}, B.~C. 2018, \araa, 56, 541

\bibitem[{{Hughes} {et~al.}(2017){Hughes}, {Lieman-Sifry}, {Flaherty}, {Daley},
  {Roberge}, {K{\'o}sp{\'a}l}, {Mo{\'o}r}, {Kamp}, {Wilner}, {Andrews},
  {Kastner}, \& {{\'A}brah{\'a}m}}]{2017ApJ...839...86H}
{Hughes}, A.~M., {Lieman-Sifry}, J., {Flaherty}, K.~M., {et~al.} 2017, \apj,
  839, 86

\bibitem[{{Kral} {et~al.}(2017){Kral}, {Matr{\`a}}, {Wyatt}, \&
  {Kennedy}}]{2017MNRAS.469..521K}
{Kral}, Q., {Matr{\`a}}, L., {Wyatt}, M.~C., \& {Kennedy}, G.~M. 2017, \mnras,
  469, 521

\bibitem[{{Kral} {et~al.}(2016){Kral}, {Wyatt}, {Carswell}, {Pringle},
  {Matr{\`a}}, \& {Juh{\'a}sz}}]{2016MNRAS.461..845K}
{Kral}, Q., {Wyatt}, M., {Carswell}, R.~F., {et~al.} 2016, \mnras, 461, 845

\bibitem[{{Marino} {et~al.}(2020){Marino}, {Flock}, {Henning}, {Kral},
  {Matr{\`a}}, \& {Wyatt}}]{2020MNRAS.492.4409M}
{Marino}, S., {Flock}, M., {Henning}, T., {et~al.} 2020, \mnras, 492, 4409

\bibitem[{{Marschall} {et~al.}(2020){Marschall}, {Markkanen}, {Gerig},
  {Pinz{\'o}n-Rodr{\'\i}guez}, {Thomas}, \& {Wu}}]{2020FrP.....8..227M}
{Marschall}, R., {Markkanen}, J., {Gerig}, S.-B., {et~al.} 2020, Frontiers in
  Physics, 8, 227

\bibitem[{{Miotello} {et~al.}(2017){Miotello}, {van Dishoeck}, {Williams},
  {Ansdell}, {Guidi}, {Hogerheijde}, {Manara}, {Tazzari}, {Testi}, {van der
  Marel}, \& {van Terwisga}}]{2017A&A...599A.113M}
{Miotello}, A., {van Dishoeck}, E.~F., {Williams}, J.~P., {et~al.} 2017, \aap,
  599, A113

\bibitem[{{Mo{\'o}r} {et~al.}(2017){Mo{\'o}r}, {Cur{\'e}}, {K{\'o}sp{\'a}l},
  {{\'A}brah{\'a}m}, {Csengeri}, {Eiroa}, {Gunawan}, {Henning}, {Hughes},
  {Juh{\'a}sz}, {Pawellek}, \& {Wyatt}}]{2017ApJ...849..123M}
{Mo{\'o}r}, A., {Cur{\'e}}, M., {K{\'o}sp{\'a}l}, {\'A}., {et~al.} 2017, \apj,
  849, 123

\bibitem[{{Mo{\'o}r} {et~al.}(2019){Mo{\'o}r}, {Kral}, {{\'A}brah{\'a}m},
  {K{\'o}sp{\'a}l}, {Dutrey}, {Di Folco}, {Hughes}, {Juh{\'a}sz}, {Pascucci},
  \& {Pawellek}}]{2019ApJ...884..108M}
{Mo{\'o}r}, A., {Kral}, Q., {{\'A}brah{\'a}m}, P., {et~al.} 2019, \apj, 884,
  108

\bibitem[{{Mumma} \& {Charnley}(2011)}]{2011ARA&A..49..471M}
{Mumma}, M.~J. \& {Charnley}, S.~B. 2011, \araa, 49, 471

\bibitem[{{Nakatani} {et~al.}(2021){Nakatani}, {Kobayashi}, {Kuiper}, {Nomura},
  \& {Aikawa}}]{2021ApJ...915...90N}
{Nakatani}, R., {Kobayashi}, H., {Kuiper}, R., {Nomura}, H., \& {Aikawa}, Y.
  2021, \apj, 915, 90

\bibitem[{{Nakatani} {et~al.}(2023){Nakatani}, {Turner}, {Hasegawa}, {Cataldi},
  {Aikawa}, {Marino}, \& {Kobayashi}}]{2023ApJ...959L..28N}
{Nakatani}, R., {Turner}, N.~J., {Hasegawa}, Y., {et~al.} 2023, \apjl, 959, L28

\bibitem[{{{\"O}berg} {et~al.}(2009){{\"O}berg}, {Linnartz}, {Visser}, \& {van
  Dishoeck}}]{2009ApJ...693.1209O}
{{\"O}berg}, K.~I., {Linnartz}, H., {Visser}, R., \& {van Dishoeck}, E.~F.
  2009, \apj, 693, 1209

\bibitem[{{Rebollido} {et~al.}(2022){Rebollido}, {Ribas}, {de
  Gregorio-Monsalvo}, {Villaver}, {Montesinos}, {Chen}, {Canovas}, {Henning},
  {Mo{\'o}r}, {Perrin}, {Rivi{\`e}re-Marichalar}, \&
  {Eiroa}}]{2022MNRAS.509..693R}
{Rebollido}, I., {Ribas}, {\'A}., {de Gregorio-Monsalvo}, I., {et~al.} 2022,
  \mnras, 509, 693

\bibitem[{{Slettebak}(1975)}]{1975ApJ...197..137S}
{Slettebak}, A. 1975, \apj, 197, 137

\bibitem[{{Smirnov-Pinchukov} {et~al.}(2022){Smirnov-Pinchukov}, {Mo{\'o}r},
  {Semenov}, {{\'A}brah{\'a}m}, {Henning}, {K{\'o}sp{\'a}l}, {Hughes}, \& {di
  Folco}}]{2022MNRAS.510.1148S}
{Smirnov-Pinchukov}, G.~V., {Mo{\'o}r}, A., {Semenov}, D.~A., {et~al.} 2022,
  \mnras, 510, 1148

\bibitem[{{Tabone} {et~al.}(2021){Tabone}, {van Hemert}, {van Dishoeck}, \&
  {Black}}]{2021A&A...650A.192T}
{Tabone}, B., {van Hemert}, M.~C., {van Dishoeck}, E.~F., \& {Black}, J.~H.
  2021, \aap, 650, A192

\bibitem[{{Trapman} {et~al.}(2022){Trapman}, {Zhang}, {van't Hoff},
  {Hogerheijde}, \& {Bergin}}]{2022ApJ...926L...2T}
{Trapman}, L., {Zhang}, K., {van't Hoff}, M. L.~R., {Hogerheijde}, M.~R., \&
  {Bergin}, E.~A. 2022, \apjl, 926, L2

\bibitem[{{van Dishoeck} {et~al.}(2006){van Dishoeck}, {Jonkheid}, \& {van
  Hemert}}]{2006FaDi..133..231V}
{van Dishoeck}, E.~F., {Jonkheid}, B., \& {van Hemert}, M.~C. 2006, Faraday
  Discussions, 133, 231

\bibitem[{{Westley} {et~al.}(1995){Westley}, {Baragiola}, {Johnson}, \&
  {Baratta}}]{1995Natur.373..405W}
{Westley}, M.~S., {Baragiola}, R.~A., {Johnson}, R.~E., \& {Baratta}, G.~A.
  1995, \nat, 373, 405

\bibitem[{{Willacy} \& {Langer}(2000)}]{2000ApJ...544..903W}
{Willacy}, K. \& {Langer}, W.~D. 2000, \apj, 544, 903

\bibitem[{{Wyatt}(2008)}]{2008ARA&A..46..339W}
{Wyatt}, M.~C. 2008, \araa, 46, 339

\bibitem[{{Zannese} {et~al.}(2024){Zannese}, {Tabone}, {Habart}, {Goicoechea},
  {Zanchet}, {van Dishoeck}, {van Hemert}, {Black}, {Tielens}, {Veselinova},
  {Jambrina}, {Menendez}, {Verdasco}, {Aoiz}, {Gonzalez-Sanchez}, {Trahin},
  {Dartois}, {Bern{\'e}}, {Peeters}, {He}, {Sidhu}, {Chown}, {Schroetter}, {Van
  De Putte}, {Canin}, {Alarc{\'o}n}, {Abergel}, {Bergin}, {Bernard-Salas},
  {Boersma}, {Bron}, {Cami}, {Dicken}, {Elyajouri}, {Fuente}, {Gordon}, {Issa},
  {Joblin}, {Kannavou}, {Khan}, {Languignon}, {Le Gal}, {Maragkoudakis},
  {Meshaka}, {Okada}, {Onaka}, {Pasquini}, {Pound}, {Robberto}, {R{\"o}llig},
  {Schefter}, {Schirmer}, {Vicente}, \& {Wolfire}}]{2024NatAs...8..577Z}
{Zannese}, M., {Tabone}, B., {Habart}, E., {et~al.} 2024, Nature Astronomy, 8,
  577

\bibitem[{{Zuckerman} \& {Song}(2012)}]{2012ApJ...758...77Z}
{Zuckerman}, B. \& {Song}, I. 2012, \apj, 758, 77

\end{thebibliography}

\begin{appendix}

\section{Production and destruction timescales of water vapor in debris disk-like environments} \label{sec:app_timescale}

Here, we briefly discuss production and destruction timescales of water vapor in debris disk-like environments.
We adopt typical values of debris disk parameters to compute these timescales; 
a disk is located $R \sim 100$ au away from the host star,
its vertical thickness is $h \sim 6 $ \%, and its radial extent is $\Delta R \sim 50$ au \citep[e.g.,][]{2018ARA&A..56..541H}.
We show below that the assumption adopted in this work likely holds for such a typical debris disk,
that is, in the debris disk-like environment, the water vapor abundance is regulated by competition between photodesorption and photodissociation.
Most quantities are defined in the main text.

Water vapor is produced by both vaporization of water ice by destructive collisions and photodesorption.
In Section \ref{sec:water}, detailed collision properties and the resulting production rate are not considered.
This is motivated by the assumption that the steady state collisional cascade might keep the water vapor abundance sufficiently high,
so that detailed collision properties might not affect the abundance considerably.
In contrast to protoplanetary disks, however, water vapor may be produced by collisions between small dust grains in debris disks.
We estimate the corresponding timescale and compare it with the timescale of photodesorption.

The collision timescale between dust grains is estimated as
\begin{equation}
\tau_{\rm coll} \sim \frac{1}{n_{\rm dust} \sigma_{\rm dust} v_{\rm rel}},
\end{equation}
where $n_{\rm dust}= N_{\rm dust}/ ( 2 \pi R^2 \Delta R h)$ is the number density of dust particles distributed in debris disks,
$v_{\rm rel}= e v_{\rm Kep}$ is the relative velocity between the particles, 
$e \sim 2h$ is the corresponding eccentricity, and $v_{\rm Kep}$ is the Keplerian velocity at $R$.
Adopting the typical disk parameters described above, $\tau_{\rm coll}$ becomes
\begin{eqnarray}
\tau_{\rm coll}  & \simeq & 3.7 \times 10^2 \mbox{ yr} \left( \frac{ R }{ 100 \mbox{ au} } \right)^{5/2} \left( \frac{ \Delta R }{ 50 \mbox{ au} } \right)  \\ \nonumber
                       & \times   & \left( \frac{ M_{Z, \rm ref} }{ 10^{-1} M_{\oplus}} \right)^{-1}  \left( \frac{ r_{\rm dust} }{ 1~ \mu\mbox{m} } \right),
\end{eqnarray}
where $ M_{Z, \rm ref}$ is estimated at $\overline{M}_{\rm CO, tran}$ (Figure \ref{fig2}).

If the water vapor abundance is regulated by competition between photodesorption and photodissociation as assumed in Section \ref{sec:water},
then the timescale of photodesorption ($\tau_{\rm deso}$) becomes comparable to that of photodissociation ($\tau_{\rm diss}$), that is,
\begin{equation}
\label{eq:app_lifetime}
\tau_{\rm deso} \sim \tau_{\rm diss} \sim \frac{1}{R_0} \sim 3.2 \times 10 \mbox{ yr}.
\end{equation}
This is a conservative upper limit as interstellar UV radiation is considered (see Appendix \ref{sec:app_UV}).

Thus, we find $\tau_{\rm deso} < \tau_{\rm coll}$ for typical debris disk-like environments.

On the other hand, water vapor is destroyed by both photodissociation and freezeout onto dust grains.
The freezeout timescale is estimated as
\begin{equation}
\tau_{\rm free} \sim  \frac{1}{n_{\rm dust} \sigma_{\rm dust} c_{\rm s}},
\end{equation}
where $c_{\rm s}$ is the sound speed of water vapor.
Adopting the typical disk parameters, $\tau_{\rm free}$ becomes
\begin{eqnarray}
\tau_{\rm free}  & \simeq & 2.1 \times 10^2 \mbox{ yr} \left( \frac{ R }{ 100 \mbox{ au} } \right)^{2} \left( \frac{ \Delta R }{ 50 \mbox{ au} } \right)   \\ \nonumber
                       & \times   & \left( \frac{ M_{Z, \rm ref} }{ 10^{-1} M_{\oplus}} \right)^{-1} \left( \frac{ T_{\rm gas} }{ 100 \mbox{ K} } \right)^{-1/2}  \left( \frac{ r_{\rm dust} }{ 1~ \mu\mbox{m} } \right),
\end{eqnarray}
where $ M_{Z, \rm ref}$ is again estimated at $\overline{M}_{\rm CO, tran}$ (Figure \ref{fig2}), and the water vapor temperature is assumed to be 100 K.

Thus, we find $\tau_{\rm diss} < \tau_{\rm free}$ for typical debris disk-like environments.

In summary, the assumption that the water vapor abundance is determined by 
competition between photodesorption and photodissociation likely holds for debris disks that have typical disk parameters.
It should be noted that for debris disks with $M_{Z, \rm ref} > 0.1 M_{\oplus}$ (or $\overline{M}_{\rm CO} > 10^{-3} M_{\oplus}$), 
both $\tau_{\rm coll}$ and $\tau_{\rm free}$ can become shorter than $\tau_{\rm deso}$ and $\tau_{\rm diss}$, respectively.
For this case, the assumption becomes no longer valid, and the water vapor abundance could be very low.
However, the current observations confirm that debris disks whose observed CO gas can be explained by Hypothesis 2, 
satisfy the condition that $M_{Z, \rm ref} < 0.1 M_{\oplus}$ (or $\overline{M}_{\rm CO} < 10^{-3} M_{\oplus}$).

\section{The effect of stellar UV radiation} \label{sec:app_UV}

Here, we briefly discuss the effect of stellar UV radiation on the abundance estimate and lifetime of water vapor.
We show below that our abundance estimates do not change very much even if the stellar UV radiation is included,
but target-dependent calculations are needed to reliably estimate the lifetime.

First, our calculations conducted in Section \ref{sec:water} focus on the interstellar UV radiation, 
given that the presence of CO gas is confirmed only for debris disks, 
where interstellar UV radiation plays a major role in photodissociation than stellar UV radiation \citep{2017MNRAS.469..521K}.
For this case, the abundance of cold water vapor becomes independent of the UV flux 
because the optical depth effect that is applied to both the production and destruction rates cancels out.
As a result, only the photodissociation cross-section ($\sigma_{\rm photo}$) emerges in equation (\ref{eq:balance}), 
which is a function of the UV spectrum.

Second, if stellar UV radiation dominates over interstellar UV radiation, 
then the abundance of cold water vapor again becomes insensitive to the UV flux for the same reason;
the optical depth effect cancels out.
In this case, one needs to adopt the value of $\sigma_{\rm photo}$ that is computed from stellar UV spectrum.
Young stars are known to have excess UV emission, 
and specific resonance lines like Lyman $\alpha$ dominate the spectrum \citep[e.g.,][]{2003ApJ...591L.159B}.
The value of $\sigma_{\rm photo}$ at the Lyman $\alpha$ line is $\sigma_{\rm photo}^{\rm Stellar} =1.2 \times10^{-17} \mbox{ cm}^{2}$ \citep{2006FaDi..133..231V}.
Interestingly, this value is comparable to that of the cross-section derived for interstellar UV radiation,
resulting in similar abundance estimates of cold water vapor.

Third, if the contribution of stellar UV radiation becomes comparable to that of interstellar UV radiation,
then it is possible to mathematically assume that the resulting contribution would be twice the former one (or the latter one).
The resulting abundance estimate is a factor of two smaller than the case where either interstellar UV radiation or stellar UV radiation dominates.

It thus can be concluded that our abundance estimates do not change very much even if stellar UV radiation is taken into account.

On the other hand, the lifetime of cold water vapor depends on the UV radiation source.
Given that most debris disks would be optically thin at a wide range of wavelengths,
it would be reasonable to consider that the lifetime derived from unshielded interstellar UV radiation gives an upper limit,
which is about 30 yrs (see equation (\ref{eq:app_lifetime})).
If stellar UV radiation dominates over interstellar UV radiation,
then target-dependent calculations are needed 
because the photodissociation rate becomes functions of both the position of debris disks and the optical depth computed along from the host star to the disks.
The resulting lifetime could be much shorter; 
for instance, a previous study estimates the lifetime of water vapor to be only 3.5 days for the debris disk around $\beta$ Pic \citep{2019A&A...628A.127C}.
For such targets, detecting cold water vapor would be challenging.

In summary, our abundance estimates would be reasonable for both interstellar and stellar UV radiation,
while feasibility of observing cold water vapor in debris disks depends on the UV radiation source;
if stellar UV radiation dominates, then the detection of the vapor may be hard due to its short lifetime.

\section{Monte-Carlo-based realization of the CO and water gas masses} \label{sec:app_mc}

Here, we briefly describe how Monte-Carlo-based realization of the CO and water gas masses is conducted.

As described in Section \ref{sec:water}, 
it is key to statistically identify a non-monotonic behavior of the water gas mass as a function of CO gas mass.
On the left panel of Figure \ref{fig3}, constant values of the conversion factor 
(i.e., equations (\ref{eq:M_H2O_CO_hyp1}) and (\ref{eq:M_H2O_CO_hyp2}) for Hypotheses 1 and 2, respectively) are used.
One may then wonder how many targets with what mass sensitivity of these gas species would be needed for statistical identification, 
if the values vary.
This can be addressed by performing Monte-Carlo-based realization of the CO and water gas masses.

We adopt a simplified approach, 
wherein the CO gas mass of targets is assumed to be logarithmically uniform in the observed range (i.e., $10^{-1} - 4 \times 10^{-7} M_{\oplus}$)
and the conversion factor is assumed to be log normal
with the peak value described in equations (\ref{eq:M_H2O_CO_hyp1}) and (\ref{eq:M_H2O_CO_hyp2}) for Hypotheses 1 and 2, respectively.
We have confirmed that the resulting log normal distribution covers the mass ratio of water to CO gas found in our order-of-magnitude analysis (Table \ref{table1}).
Also, we assume that switching occurs at the CO gas mass of $10^{-3} M_{\oplus}$;
we have adopted smooth transitions with the transition width of up to $5 \times 10^{-4} M_{\oplus}$ and confirmed that the results do not change very much.

\begin{figure}
\begin{center}
\includegraphics[width=8cm]{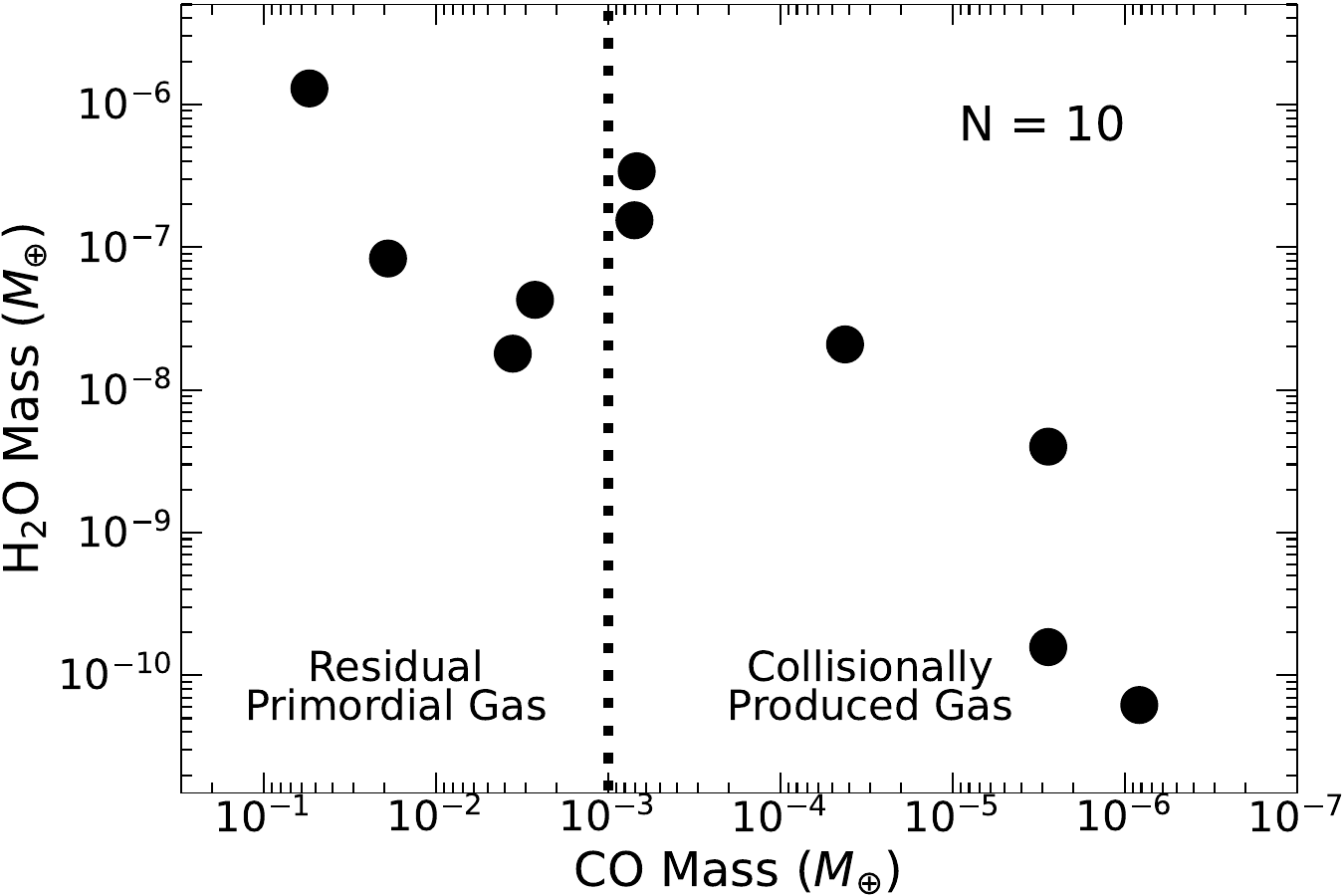}
\includegraphics[width=8cm]{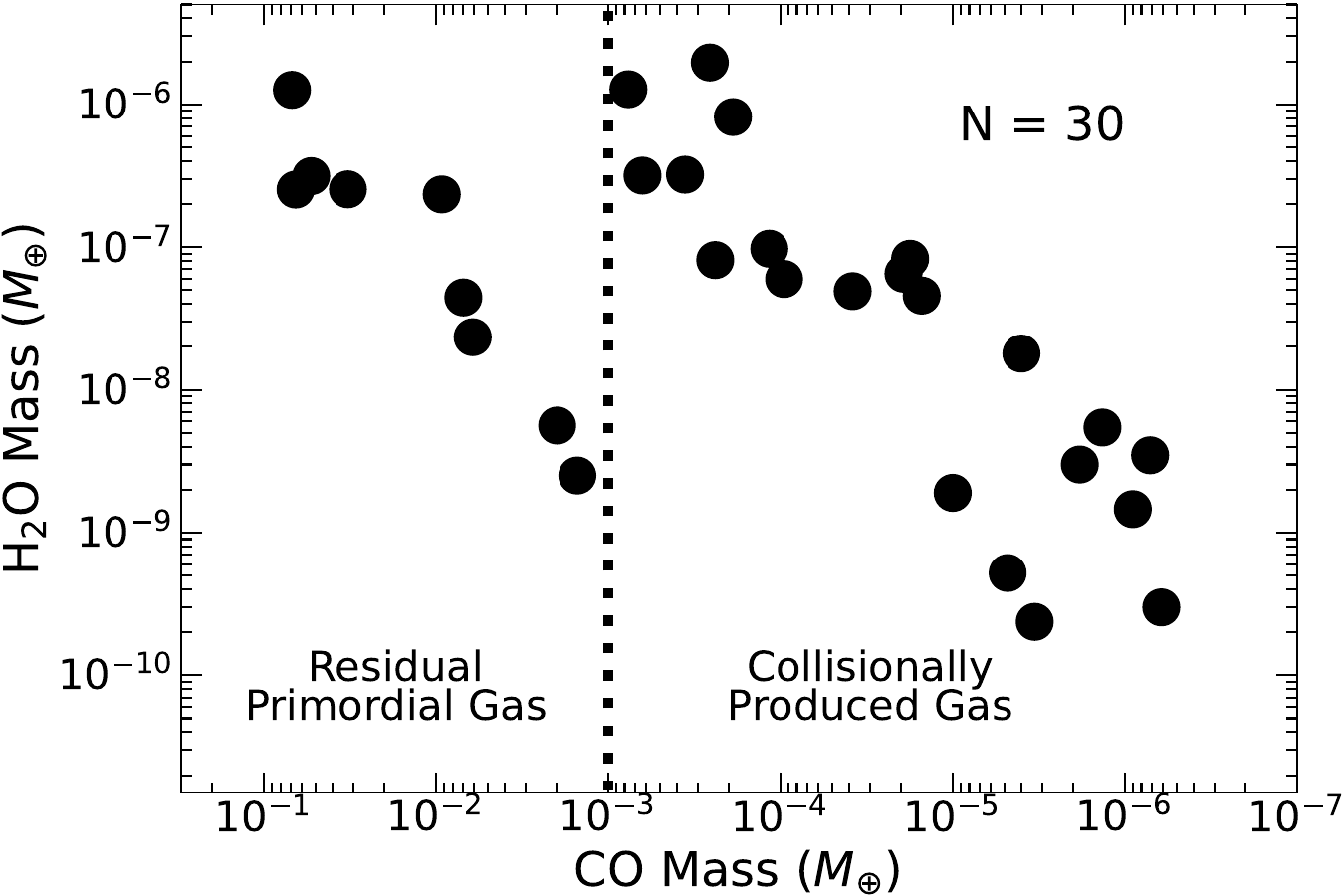}
\caption{Realization of the CO and water gas masses by the Monte-Carlo approach.
On the top panel, 10 realizations are done (i.e., 10 targets), while on the bottom panel, 30 realizations are performed.
Statistical identification of the non-monotonic behavior is very likely possible if $N \gtrsim 10$ 
and if the mass sensitivities of CO gas and water gas are an order of $10^{-4} M_{\oplus}$ and of $10^{-9}-10^{-8} M_{\oplus}$ or lower, respectively.}
\label{fig5}
\end{center}
\end{figure}

Figure \ref{fig5} shows our results.
We consider two sample sizes: one is $N=10$, and the other is $N=30$.
It should be noted that the total number of observed systems is about 20 currently.
First, we confirm by eye that statistical identification of the non-monotonic behavior is very likely possible if $N \ga 10$.
Second, given that statistical identification becomes reliable only if error bars of data points do not overlap,
the mass sensitivity of CO gas should be an order of $10^{-4} M_{\oplus}$ or lower for both sample sizes;
equivalently, errors should be a factor of few (or more) smaller than measurement values.
For water vapor, the mass sensitivity should be $10^{-9} M_{\oplus}$ or lower for $N=10$ and $10^{-8} M_{\oplus}$ or lower for $N=30$.

In actual observations, the gas mass is computed from flux measurements.
Observed flux is a function of targets' distance.
As described in Section \ref{sec:disc}, CO-gas rich debris disks are $>100$ pc away from the Earth.
Therefore, a desired sensitivity improvement factor over {\it Herschel} would be 100 or more.

\end{appendix}

\end{document}